\documentclass[12pt,preprint]{aastex}
\begin{document}

\begin{center}
{\LARGE Multicomponent theory of buoyancy instabilities in magnetized
astrophysical plasmas: MHD analysis revisited}

\bigskip

{\large Anatoly K. Nekrasov}

Institute of Physics of the Earth, Russian Academy of Sciences, 123995
Moscow, Russia

anatoli.nekrassov@t-online.de

\bigskip

{\large and}

\bigskip

{\large Mohsen Shadmehri}

Department of Mathematical Physics, National University of Ireland Maynooth,
Maynooth, Co. Kildare, Ireland

mshadmehri@thphys.nuim.ie

\bigskip

{\large ABSTRACT}
\end{center}

We develop a theory of buoyancy instabilities of the electron-ion plasma
with the heat flux based on not the MHD equations, but using the
multicomponent plasma approach. We investigate a geometry in which the
background magnetic field, gravity, and stratification are directed along
one axis. No simplifications usual for the MHD-approach in studying these
instabilities are used. The background electron thermal flux and collisions
between electrons and ions are included. We derive the simple dispersion
relation, which shows that the thermal flux perturbation generally
stabilizes an instability. There is a narrow region of the temperature
gradient, where an instability is possible. This result contradicts to a
conclusion obtained in the MHD-approach. We show that the reason of this
contradiction is the simplified assumptions used in the MHD analysis of
buoyancy instabilities and the role of the longitudinal electric field
perturbation, which is not captured by the MHD equations. Our dispersion
relation also shows that a medium with the electron thermal flux can be
unstable, if the temperature gradient of ions and electrons have the
opposite signs. The results obtained can be applied to ICM and clusters of
galaxies.

\bigskip \textit{Subject headings: }convection - instabilities - magnetic
fields - plasmas - waves

\bigskip

\section{INTRODUCTION}

Various instability mechanisms are studied to understand some of the main
features and processes in the astrophysical objects depending on their
physical properties. Convective or buoyancy instability arising as a result
of stratification is among those instabilities that may operate under
different circumstances from the stellar interiors (e.g., Schwarzschild
1958), accretion disks (Balbus 2000, 2001), and neutron stars (Chang \&
Quataert 2009) to the hot accretion flows (e.g., Narayan et al. 2000, 2002)
and even galaxy clusters and intercluster medium (ICM) (e.g., Quataert 2008;
Sharma et al. 2009; Ren et al. 2009). Analogous instabilities also exist in
the neutral atmosphere of the Earth and ocean (Gossard \& Hooke 1975;
Pedlosky 1982). Diversity of the astrophysical objects, in which convective
instabilities may have a significant role, leading to turbulence and
anomalous energy and matter transport, is a good motivation to explore this
instability either through linear analytical analysis or by direct numerical
simulations from different physical point of views. Although the significant
role of convection in the transport of energy in stellar interiors is a
well-known physical process, theoretical efforts to understand convective
energy transport in the tenuous and hot plasmas such as ICM (Sarazin 1988)
have lead to some results over recent years.

According to the standard Schwarzschild criterion, a thermally stratified
fluid is convectively unstable when the entropy increases in the direction
of gravity (Schwarzschild 1958). By taking into account the anisotropic heat
flux in plasmas where the mean free path of ions and electrons is much
larger than their Larmor radius, one obtains additional instabilities for
short wave numbers with larger growth rates than that without thermal flux.
These instabilities have been shown to arise when the temperature increases
in the direction of gravity at the absence of the background thermal flux
(the magnetothermal instability (MTI)) (Balbus 2000, 2001) and when the
temperature decreases along gravity at the presence of the latter (the heat
buoyancy instability (HBI)) (Quataert 2008). Both MTI and HBI have been
simulated in 2D and 3D by many authors over recent years (e.g., Parrish et
al. 2008; Parrish \& Quataert 2008; Parrish et al. 2009). Following recent
achievements in the convective theory, it has attracted attention of the
authors for analyzing its possible role in ICM after a long time
discounting. Majority of the mass of a cluster of galaxies is in the dark
matter. However, around 1/6 of its mass consists of a hot, magnetized, and
low density plasma known as ICM. The electron density is $n_{e}\simeq 10^{-2}
$ to $10^{-1}$ cm$^{-3}$ at the central parts of ICM. The electron
temperature $T_{e}$ is measured of the order of a few keV, though the ion
temperature $T_{i}$ has not yet been measured directly (e.g., Fabian et al.
2006; Sanders et al. 2010). The magnetic field strength $B$ in ICM is
estimated to be in the range 0.1-10 $\mu $G depending on where the
measurement is made (Carilli \& Taylor 2002) which implies a dynamically
weak magnetic field with $\beta =8\pi n_{e}T_{e}/B^{2}\approx 200-2000$.
Thus, ICM with the ion Larmor radius $10^{8-9}$ cm ($T_{i}\sim T_{e}$) and
the mean free path $10^{22-23}$ cm is classified as a weakly collisional
plasma (Carilli \& Taylor 2002). In simulating ICM, it is important to
consider anisotropic viscosity as well because the Reynolds number is very
low (Lyutikov 2007, 2008). Another important physical agent is cosmic rays.
Recent studies show that centrally concentrated cosmic rays have a
destabilizing effect on the convection in ICM (Chandran \& Dennis 2006;
Rasera and Chandran 2008).

Theoretical models applied for study of buoyancy instabilities are based on
the ideal magnetohydrodynamic (MHD) equations (Balbus 2000, 2001; Quataert
2008, Chang \& Quataert 2009; Ren et al. 2009). Using of these equations
permits us comparatively easily to consider different problems. However, the
ideal MHD does not capture some important effects. One of the such effects
is the nonzero longitudinal electric field perturbation along the
background magnetic field. As we show here, the contribution of currents due
to this small field to the dispersion relation can be of the same order of
magnitude as that due to other electric field components. Besides, the MHD
equations do not take into account the very existence of various charged and
neutral species with different masses and electric charges and their
collisions between each others and therefore can not be applied to
multicomponent systems. On the contrary, the plasma $\mathbf{E}$-approach
deals with dynamical equations for each species. From Faraday's and Ampere's
laws one obtains equations for the electric field components. Such an
approach allows us to follow the movement and changing of parameters of each
species separately and obtain rigorous conditions of consideration and
physical consequences in specific cases. This approach permits us to include
various species of ions and dust grains having different charges and masses.
In this way, streaming instabilities of rotating multicomponent objects
(accretion disks, molecular clouds and so on) have been investigated by
Nekrasov (e.g., 2008, 2009 a, 2009 b), which have growth rates much larger
than that of the magnetorotational instability (Balbus 1991). In some cases,
the standard methods used in MHD leads to conclusions that are different
from those obtained by the method using the electric field perturbations.
One of a such example is considered in Nekrasov (2009 c).

In this paper, we apply a multicomponent approach to study buoyancy
instabilities in magnetized electron-ion astrophysical plasmas with the
background electron thermal flux. We include collisions between electrons
and ions. However, we adopt here that cyclotron frequencies of species are
much larger than their collision frequencies. Such conditions are typical
for ICM and galaxy clusters. In this case, as it is known, the heat flux is
anisotropic and directed along the magnetic field lines (Braginskii 1965).
We consider a geometry in which gravity, stratification, and the background
magnetic field are all directed along one ($z$-) axis. In our approach, it
is important to obtain exact expressions for species' velocities in an
inhomogeneous medium. We give main equations and results. However, for those
who are not interested in the mathematical details, they can directly refer
to Sections 7 and 8. The dispersion relation is obtained for cases, in which
the background heat flux is absent or present. This gives a possibility to
compare two cases. Solutions of the dispersion relation are discussed.

The paper is organized as follows. In Section 2, the fundamental equations
are given. An equilibrium state is considered in Section 3. Perturbed ion
velocity, number density, and thermal pressure are obtained in Section 4. In
Section 5, we consider the perturbed velocity and temperature for electrons.
Components of the dielectric permeability tensor are found in Section 6.

\bigskip

\section{BASIC EQUATIONS}

We start with the following equations for ions:
\begin{equation}
\frac{\partial \mathbf{v}_{i}}{\partial t}\mathbf{=-}\frac{\mathbf{\nabla }%
p_{i}}{m_{i}n_{i}}+\mathbf{g+}\frac{q_{i}}{m_{i}}\mathbf{E}+\frac{q_{i}}{%
m_{i}c}\mathbf{v}_{i}\times \mathbf{B}-\nu _{ie}\left( \mathbf{v}_{i}-%
\mathbf{v}_{e}\right) ,
\end{equation}
the momentum equation,
\begin{equation}
\frac{\partial n_{i}}{\partial t}+\mathbf{\nabla }\cdot n_{i}\mathbf{v}%
_{i}=0,
\end{equation}
the continuity equation, and
\begin{equation}
\frac{\partial p_{i}}{\partial t}+\mathbf{v}_{i}\cdot \mathbf{\nabla }%
p_{i}+\gamma p_{i}\mathbf{\nabla }\cdot \mathbf{v}_{i}=0,
\end{equation}
the pressure equation. The corresponding equations for electrons are:
\begin{equation}
\mathbf{0=-}\frac{\mathbf{\nabla }p_{e}}{n_{e}}+q_{e}\mathbf{E}+\frac{q_{e}}{%
c}\mathbf{v}_{e}\times \mathbf{B}-m_{e}\nu _{ei}\left( \mathbf{v}_{e}-%
\mathbf{v}_{i}\right) ,
\end{equation}%
\begin{equation}
\frac{\partial n_{e}}{\partial t}+\mathbf{\nabla }\cdot n_{e}\mathbf{v}%
_{e}=0,
\end{equation}%
\begin{equation}
\frac{\partial p_{e}}{\partial t}+\mathbf{v}_{e}\cdot \mathbf{\nabla }%
p_{e}+\gamma p_{e}\mathbf{\nabla }\cdot \mathbf{v}_{e}=\lambda -\left(
\gamma -1\right) \mathbf{\nabla \cdot q}_{e},
\end{equation}%
\begin{equation}
\frac{\partial T_{e}}{\partial t}+\mathbf{v}_{e}\cdot \mathbf{\nabla }%
T_{e}+\left( \gamma -1\right) T_{e}\mathbf{\nabla }\cdot \mathbf{v}_{e}=%
\frac{\lambda }{n_{e}}-\left( \gamma -1\right) \frac{1}{n_{e}}\mathbf{\nabla
\cdot q}_{e},
\end{equation}
the temperature equation, where $\mathbf{q}_{e}$ is the electron heat flux
(Braginskii 1965). We neglect inertia of the electrons. In Equations
(1)-(7), $q_{j}$ and $m_{j}$ are the charge and mass of species $j=i,e$, $%
\mathbf{v}_{j}$ is the hydrodynamic velocity, $n_{j}$ is the number density,
$p_{j}=n_{j}T_{j}$ is the thermal pressure, $T_{j}$ is the temperature, $\nu
_{ie}$ ($\nu _{ei}$) is the collision frequency of ions (electrons) with
electrons (ions), $\mathbf{g}$ is gravity, $\mathbf{E}$\textbf{\ }and $%
\mathbf{B}$ are the electric and magnetic fields, $c$ is the speed of light
in vacuum, and $\gamma $ is the adiabatic constant. We assume the electrons
to be magnetized when their cyclotron frequency $\omega
_{ce}=q_{e}B/m_{e}c\gg \nu _{ee}$, where $\nu _{ee}$ is the the
electron-electron collision frequency. In this case, the electron thermal
flux is mainly directed along the magnetic field,%
\begin{equation}
\mathbf{q}_{e}=-\chi _{e}\mathbf{b}\left( \mathbf{b\cdot \nabla }\right)
T_{e},
\end{equation}
where $\chi _{e}$ is the electron thermal conductivity coefficient and $%
\mathbf{b=B/}B$ is the unit vector along the magnetic field (Braginskii
1965). The term $\lambda $ compensates the temperature change as a result of
the equilibrium heat flux. We take only into account the electron thermal
conductivity by equation (8), because the corresponding ion conductivity is
considerably smaller (Braginskii 1965).

Electromagnetic equations are Faraday's law
\begin{equation}
\mathbf{\nabla \times E=-}\frac{1}{c}\frac{\partial \mathbf{B}}{\partial t}
\end{equation}
and Ampere`s law
\begin{equation}
\mathbf{\nabla \times B=}\frac{4\pi }{c}\mathbf{j,}
\end{equation}
where $\mathbf{j=}\sum_{j}q_{j}n_{j}\mathbf{v}_{j}.$ We consider the wave
processes with typical time-scales much larger than the time the light
spends to cover the wavelength of perturbations. In this case, one can
neglect the displacement current in Equation (10) that results in
quasineutrality both in electromagnetic and purely electrostatic
perturbations. The magnetic field $\mathbf{B}$ includes the background
magnetic field $\mathbf{B}_{0}$, the magnetic field $\mathbf{B}_{0cur}$ of
the background current (when it presents), and the perturbed magnetic field.

\bigskip

\section{EQUILIBRIUM STATE}

At first, we consider an equilibrium state. We assume that background
velocities are absent. In this paper, we study configuration, in which the
background magnetic field, gravity, and stratification are directed along
the $z$-axis. Let, for definiteness, $\mathbf{g}$ be $\mathbf{g=-z}g$, where
$g>0$ and $\mathbf{z}$ is the unit vector along the $z$-direction. Then,
Equations (1) and (4) give
\begin{equation}
g_{i}=-\frac{1}{m_{i}n_{i0}}\frac{\partial p_{i0}}{\partial z}=g-\frac{q_{i}%
}{m_{i}}E_{0},
\end{equation}%
\begin{equation}
g_{e}=-\frac{1}{m_{i}n_{e0}}\frac{\partial p_{e0}}{\partial z}=\frac{q_{i}}{%
m_{i}}E_{0},
\end{equation}
where (and below) the index $0$ denotes equilibrium values. Here and below
we assume that $q_{i}=-q_{e}$. We see that equilibrium distributions of ions
and electrons influence each other through the background electric field $%
E_{0}$. In the case $n_{i0}=n_{e0}$ ( this equality is satisfied for the
two-component plasma) and $T_{i0}=T_{e0}$, we obtain $g_{i}=g_{e}=$ $g/2$.
Thus, we have $E_{0}=m_{i}g/2q_{i}$. The presence of the third component,
for example, of the cold dust grains with the charge $q_{d}$ and mass $%
m_{d}\gg m_{i}$ results in other value of $E_{0}=m_{d}g/q_{d}$. In this
case, the ions and electrons are in equilibrium under the action of the
thermal pressure and equilibrium electric field, being $g_{i}\simeq -g_{e}$.

\bigskip

\section{LINEAR\ ION\ PERTURBATIONS}

Let us write Equations (1)-(3) for ions in the linear approximation,%
\begin{equation}
\frac{\partial \mathbf{v}_{i1}}{\partial t}\mathbf{=-}\frac{\mathbf{\nabla }%
p_{i1}}{m_{i}n_{i0}}+\frac{\mathbf{\nabla }p_{i0}}{m_{i}n_{i0}}\frac{n_{i1}}{%
n_{i0}}+\mathbf{F}_{i1}+\frac{q_{i}}{m_{i}c}\mathbf{v}_{i1}\times \mathbf{B}%
_{0},
\end{equation}%
\begin{equation}
\frac{\partial n_{i1}}{\partial t}+v_{i1z}\frac{\partial n_{i0}}{\partial z}%
+n_{i0}\mathbf{\nabla }\cdot \mathbf{v}_{i1}=0,
\end{equation}%
\begin{equation}
\frac{\partial p_{i1}}{\partial t}+v_{i1z}\frac{\partial p_{i0}}{\partial z}%
+\gamma p_{i0}\mathbf{\nabla }\cdot \mathbf{v}_{i1}=0,
\end{equation}
where
\begin{equation}
\mathbf{F}_{i1}=\frac{q_{i}}{m_{i}}\mathbf{E}_{1}-\nu _{ie}\left( \mathbf{v}%
_{i1}-\mathbf{v}_{e1}\right),
\end{equation}
and the index $1$ denotes the perturbed variables. Below, we solve these
equations to find the perturbed velocity of ions in an inhomogeneous medium.

\subsection{Perturbed velocity of ions}

Applying the operator $\partial /\partial t$ to Equation (13) and using
Equations (14) and (15), we obtain%
\begin{equation}
\frac{\partial ^{2}\mathbf{v}_{i1}}{\partial t^{2}}\mathbf{=-}g_{i}\mathbf{%
\nabla }v_{i1z}+\frac{1}{m_{i}n_{i0}}\left[ \left( \gamma -1\right) \left(
\mathbf{\nabla }p_{i0}\right) +\gamma p_{i0}\mathbf{\nabla }\right] \mathbf{%
\nabla }\cdot \mathbf{v}_{i1}+\frac{\partial \mathbf{F}_{i1}}{\partial t}+%
\frac{q_{i}}{m_{i}c}\frac{\partial \mathbf{v}_{i1}}{\partial t}\times
\mathbf{B}_{0}.
\end{equation}
We can find solutions for the components of $\mathbf{v}_{i1}$. For
simplicity, we assume that $\partial /\partial x=0$, because a system is
symmetric in the transverse direction relative to the $z$-axis. The $x$%
-component of Equation (17) has the form
\begin{equation}
\frac{\partial v_{i1x}}{\partial t}\mathbf{=}F_{i1x}+\omega _{ci}v_{i1y},
\end{equation}
where $\omega _{ci}=q_{i}B_{0}/m_{i}c$ is the ion cyclotron frequency. For
the $y$-component of Equation (17), we obtain:%
\begin{equation}
\frac{\partial ^{2}v_{i1y}}{\partial t^{2}}\mathbf{=-}g_{i}\frac{\partial
v_{i1z}}{\partial y}+c_{si}^{2}\frac{\partial }{\partial y}\mathbf{\nabla }%
\cdot \mathbf{v}_{i1}+\frac{\partial F_{i1y}}{\partial t}-\omega _{ci}\frac{%
\partial v_{i1x}}{\partial t}.
\end{equation}%
Here, $c_{si}=\left( \gamma T_{i0}/m_{i}\right) ^{1/2}$ is the ion sound
velocity.

Using Equation (18), a relation for $v_{i1y}$ is given from Equation (19) as
follows
\begin{equation}
\left( \frac{\partial ^{2}}{\partial t^{2}}+\omega _{ci}^{2}\right)
v_{i1y}-Q_{i1y}\mathbf{=}\frac{\partial P_{i1}}{\partial y}.
\end{equation}
Then from Equation (18), we obtain%
\begin{equation}
\frac{\partial }{\omega _{ci}\partial t}\left[ \left( \frac{\partial ^{2}}{%
\partial t^{2}}+\omega _{ci}^{2}\right) v_{i1x}-Q_{i1x}\right] \mathbf{=}%
\frac{\partial P_{i1}}{\partial y}.
\end{equation}

Here, the following notations are introduced:%
\begin{equation}
P_{i1}=\mathbf{-}g_{i}v_{i1z}+c_{si}^{2}\mathbf{\nabla }\cdot \mathbf{v}%
_{i1},
\end{equation}%
\begin{equation}
Q_{i1x}=\omega _{ci}F_{i1y}+\frac{\partial F_{i1x}}{\partial t},
\end{equation}%
\begin{equation}
Q_{i1y}=-\omega _{ci}F_{i1x}+\frac{\partial F_{i1y}}{\partial t}.
\end{equation}

The value $P_{i1}$ defines the pressure perturbation (Eq. [15]). We see from
Equation (21) that when $\partial /\partial t\ll \omega _{ci}$ the thermal
pressure effect on the velocity $v_{i1x}$ is much larger than that on $%
v_{i1y}$. The $z$-component of Equation (17) takes the form
\begin{equation}
\frac{\partial }{\partial t}\left( \frac{\partial v_{i1z}}{\partial t}%
-F_{i1z}\right) \mathbf{=-}g_{i}\frac{\partial v_{i1z}}{\partial z}+\left[
\left( 1-\gamma \right) g_{i}+c_{si}^{2}\frac{\partial }{\partial z}\right]
\mathbf{\nabla }\cdot \mathbf{v}_{i1}.
\end{equation}

Let us now find $\mathbf{\nabla }\cdot \mathbf{v}_{i1}$ through $v_{i1z}$.
Differentiating Equation (20) with respect to $y$ and using expression (22),
we obtain
\begin{equation}
L_{1}\mathbf{\nabla }\cdot \mathbf{v}_{i1}\mathbf{=}L_{2}v_{i1z}+\frac{%
\partial Q_{i1y}}{\partial y},
\end{equation}
where the following operators are introduced:%
\begin{equation}
L_{1}=\frac{\partial ^{2}}{\partial t^{2}}+\omega _{ci}^{2}\mathbf{-}%
c_{si}^{2}\frac{\partial ^{2}}{\partial y^{2}},
\end{equation}%
\begin{equation}
L_{2}=\left( \frac{\partial ^{2}}{\partial t^{2}}+\omega _{ci}^{2}\right)
\frac{\partial }{\partial z}-g_{i}\frac{\partial ^{2}}{\partial y^{2}}.
\end{equation}

We can derive an equation for the longitudinal velocity $v_{i1z}$,
substituting $\mathbf{\nabla }\cdot \mathbf{v}_{i1}$ found from Equation
(26) in Equation (25),%
\begin{equation}
L_{3}v_{i1z}\mathbf{=}L_{1}\frac{\partial F_{i1z}}{\partial t}+L_{4}\frac{%
\partial Q_{i1y}}{\partial y},
\end{equation}
where operators $L_{3}$ and $L_{4}$ have the form
\begin{eqnarray}
L_{3} &=&\left( \frac{\partial ^{2}}{\partial t^{2}}+\omega _{ci}^{2}\right)
\frac{\partial ^{2}}{\partial t^{2}}-c_{si}^{2}\left( \frac{\partial ^{2}}{%
\partial y^{2}}+\frac{\partial ^{2}}{\partial z^{2}}\right) \frac{\partial
^{2}}{\partial t^{2}}-c_{si}^{2}\omega _{ci}^{2}\frac{\partial ^{2}}{%
\partial z^{2}} \\
&&+\gamma g_{i}\left( \frac{\partial ^{2}}{\partial t^{2}}+\omega
_{ci}^{2}\right) \frac{\partial }{\partial z}+c_{si}^{2}\frac{\partial L_{1}%
}{L_{1}\partial z}L_{2}+\left( 1-\gamma \right) g_{i}^{2}\frac{\partial ^{2}%
}{\partial y^{2}},  \nonumber
\end{eqnarray}

\begin{equation}
L_{4}=\left( 1-\gamma \right) g_{i}+c_{si}^{2}\left( \frac{\partial }{%
\partial z}-\frac{\partial L_{1}}{L_{1}\partial z}\right) .
\end{equation}
For obtaining expression (30), we have used expressions (27) and (28).

It is easy to see that at the absence of the background magnetic field and
without taking into account electromagnetic perturbations (the right
hand-side of Eq. [29]), the equation $L_{3}v_{i1z}=0$ describes the ion
sound and internal gravity waves. In this case, a sum of the last two terms
on the right hand-side of expression (30) is equal to $-c_{si}^{2}\omega
_{bi}^{2}\frac{\partial ^{2}}{\partial y^{2}}$, where $\omega _{bi}$ is the
(ion) Brunt-V\"{a}is\"{a}l\"{a} frequency equal to%
\begin{equation}
\omega _{bi}^{2}=\frac{g_{i}}{c_{si}^{2}}\left[ \left( \gamma -1\right)
g_{i}+\frac{\partial c_{si}^{2}}{\partial z}\right] .
\end{equation}
However, we see the existence of the background magnetic field considerably
modifies the operator $L_{3}$. Note the right hand-side of Equation (29)
describes a connection between ions and electrons through the electric field
$\mathbf{E}_{1}$ and collisions.

\subsection{Specific case for ions}

So far, we have not made any simplifications and all the equations and the
expressions are given in their general forms. Now, we consider further
perturbations with a frequency much lower than the ion cyclotron frequency
and the transverse wavelengths much larger than the ion Larmor radius. Such
conditions are typical for the astrophysical plasmas. Besides, we
investigate a part of the frequency spectrum in the region lower than the
ion sound frequency. Thus, we set%
\begin{equation}
\omega _{ci}^{2}\gg \frac{\partial ^{2}}{\partial t^{2}},c_{si}^{2}\frac{%
\partial ^{2}}{\partial y^{2}};c_{si}^{2}\frac{\partial ^{2}}{\partial z^{2}}%
\gg \frac{\partial ^{2}}{\partial t^{2}}.
\end{equation}%
In this case, operators (27), (28), (30), and (31) take the form%
\begin{eqnarray}
L_{1} &\simeq &\omega _{ci}^{2},L_{2}\simeq \omega _{ci}^{2}\frac{\partial }{%
\partial z}, \\
L_{3} &=&-\omega _{ci}^{2}\left[ \left( c_{si}^{2}\frac{\partial }{\partial z%
}-\gamma g_{i}\right) \frac{\partial }{\partial z}-\frac{\partial ^{2}}{%
\partial t^{2}}\right] ,  \nonumber \\
L_{4} &=&\left( 1-\gamma \right) g_{i}+c_{si}^{2}\frac{\partial }{\partial z}%
.  \nonumber
\end{eqnarray}%
Also, the operator $L_{3}$ can be written for a case in which
\begin{equation}
\omega _{ci}^{2}\frac{\partial ^{2}}{\partial t^{2}}\gg c_{si}^{2}\frac{%
\partial c_{si}^{2}}{\partial z}\frac{\partial ^{3}}{\partial y^{2}\partial z%
}.
\end{equation}
The small corrections in operators $L_{3}$ and $L_{4}$ are needed to be kept
because some main terms in expressions for ion and electron
velocities are equal each other (see below). Therefore, when calculating the
electric current these main terms will be canceled and small corrections to
velocities will only contribute to the current.

For the cases represented by inequalities (33) and (35) when the operators
have the form (34), the equations for $v_{i1z}$ and $\mathbf{\nabla }\cdot
\mathbf{v}_{i1}$ become%
\begin{equation}
\left[ \left( c_{si}^{2}\frac{\partial }{\partial z}-\gamma g_{i}\right)
\frac{\partial }{\partial z}-\frac{\partial ^{2}}{\partial t^{2}}\right]
v_{i1z}\mathbf{=-}\frac{\partial F_{i1z}}{\partial t}-\left[ \left( 1-\gamma
\right) g_{i}+c_{si}^{2}\frac{\partial }{\partial z}\right] \frac{\partial
Q_{i1y}}{\omega _{ci}^{2}\partial y},
\end{equation}%
\begin{equation}
\mathbf{\nabla }\cdot \mathbf{v}_{i1}\simeq \frac{\partial v_{i1z}}{\partial
z}+\frac{\partial Q_{i1y}}{\omega _{ci}^{2}\partial y}.
\end{equation}

\subsection{Ion velocity in the Fourier transform}

Calculations show that some main terms in expressions for $v_{i1z}$ (when
calculating the current), $\mathbf{\nabla }\cdot \mathbf{v}_{i1}$ and $%
P_{i1} $ are canceled. Therefore, the small terms proportional to
inhomogeneity must be taken into account. To make this correctly, we can not
make the Fourier transformation in Equations (36) and (37) to find the
perturbed ion pressure $P_{i1}$. However, firstly, we should apply the
operator $\partial /\partial z$ to this variable for using Equation (36). It
is analogous to obtain the term $\partial c_{s}^{2}/\partial z$ in
expression (32) for the Brunt-V\"{a}is\"{a}l\"{a} frequency. After that, we
can apply in a local approximation the Fourier transformation assuming the
linear perturbations to be proportional to $\exp (i\mathbf{kr-}i\omega t)$.
As a result, we obtain for the Fourier-components $v_{i1zk}$, $\mathbf{k}%
\cdot \mathbf{v}_{i1k}$ and $P_{i1k}$, where $k=\left( \mathbf{k,}\omega
\right) $, the following expressions:
\begin{equation}
v_{i1zk}=-i\frac{\omega }{k_{z}^{2}c_{si}^{2}}\left( 1-i\frac{\gamma g_{i}}{%
k_{z}c_{si}^{2}}\right) F_{i1zk}-\frac{k_{y}}{k_{z}\omega _{ci}^{2}}\left(
1-i\frac{g_{i}}{k_{z}c_{si}^{2}}\right) Q_{i1yk},
\end{equation}%
\begin{equation}
\mathbf{k}\cdot \mathbf{v}_{i1k}\mathbf{=-}i\frac{\omega }{k_{z}c_{si}^{2}}%
\left( 1-i\frac{\gamma g_{i}}{k_{z}c_{si}^{2}}\right) F_{i1zk}+i\frac{k_{y}}{%
k_{z}}\frac{g_{i}}{c_{si}^{2}\omega _{ci}^{2}}Q_{i1yk},
\end{equation}%
\begin{eqnarray}
P_{i1k} &=&\frac{\omega }{k_{z}}F_{i1zk}-i\frac{\omega }{k_{z}^{2}c_{si}^{2}}%
\left[ \left( \gamma -1\right) g_{i}+\frac{\partial c_{si}^{2}}{\partial z}%
\right] F_{i1zk} \\
&&+i\frac{k_{y}g_{i}}{k_{z}^{2}c_{si}^{2}\omega _{ci}^{2}}\left[ \left(
\gamma -1\right) g_{i}+\frac{\partial c_{si}^{2}}{\partial z}-\omega ^{2}%
\frac{c_{si}^{2}}{g_{i}}\right] Q_{i1yk}.  \nonumber
\end{eqnarray}

In expressions (38) and (39), we have omitted additional small terms at $%
Q_{i1yk}$, which are needed for calculation of $P_{i1k}$. When calculating
the current along the $z$-axis, the main term $\sim Q_{i1yk}$ in Equation
(38) will be canceled. The contribution of the first term $\sim F_{i1zk}$ to
this current has, as we shall show below, the same order of magnitude for
the buoyancy instabilities as that of the term $\sim g_{i}Q_{i1yk}$. The
same relates to expressions (39) and (40). Thus, the longitudinal electric
field perturbations must be taken into account. However, in the ideal MHD,
this field is absent. We see from expressions (38) and (39) that $\mathbf{%
\nabla }\cdot \mathbf{v}_{i1}\sim (g_{i}/c_{si}^{2})v_{i1z}$. This relation
is the same as that for the internal gravity waves in the Earth's atmosphere
(e.g., Nekrasov 1994). Using expression (40), we obtain velocities $v_{i1yk}$
and $v_{i1xk}$ from Equations (20) and (21), correspondingly.

\subsection{Perturbed ion number density and pressure}

It is followed from above that $\mathbf{\nabla }\cdot \mathbf{v}_{i1}\neq 0$%
. Let us find the perturbed ion number density and pressure in the
Fourier-representation. From Equations (14), (38) and (39), we obtain%
\begin{equation}
\frac{n_{i1k}}{n_{i0}}=-i\frac{1}{k_{z}c_{si}^{2}}F_{i1zk}-i\frac{k_{y}}{%
k_{z}c_{si}^{2}\omega \omega _{ci}^{2}}\left[ \left( \gamma -1\right) g_{i}+%
\frac{\partial c_{si}^{2}}{\partial z}\right] Q_{i1yk}.
\end{equation}
Equation (15) gives $\partial p_{i1}/\partial t=-m_{i}n_{i0}P_{i1}$. Thus,
we obtain, using Equation (40),%
\begin{equation}
\frac{p_{i1k}}{p_{i0}}=-i\frac{\gamma }{k_{z}c_{si}^{2}}F_{i1zk}+\frac{%
\gamma k_{y}g_{i}}{k_{z}^{2}c_{si}^{4}\omega \omega _{ci}^{2}}\left[ \left(
\gamma -1\right) g_{i}+\frac{\partial c_{si}^{2}}{\partial z}-\omega ^{2}%
\frac{c_{si}^{2}}{g_{i}}\right] Q_{i1yk}.
\end{equation}

Comparing Equations (41) and (42), we see that the relative perturbation of
the pressure due to the transverse electric force $Q_{i1yk}$ is much smaller
than the relative perturbation of the number density. However, these
perturbations as a result of the action of the longitudinal electric force $%
F_{i1zk}$ have the same order of magnitude. Thus, $p_{i1k}/p_{i0}$ $\sim
n_{i1k}/n_{i0}$. This result contradicts a supposition $p_{i1k}/p_{i0}$ $\ll
n_{i1k}/n_{i0}$ adopted in the MHD analysis of buoyancy instabilities
(Balbus 2000, 2001; Quataert 2008) because the latter does not take into
account the longitudinal electric field perturbations. From the results
given below, it is followed that, as we have already noted above, the both
terms on the right hand-side of Equation (41) have the same order of
magnitude.

\bigskip\

\section{LINEAR\ ELECTRON\ PERTURBATIONS}

Equations for the electrons in the linear approximation are the following:%
\begin{equation}
\mathbf{0=-}\frac{\mathbf{\nabla }p_{e1}}{n_{e0}}+\frac{\mathbf{\nabla }%
p_{e0}}{n_{e0}}\frac{n_{e1}}{n_{e0}}+\mathbf{F}_{e1}+\frac{q_{e}}{c}\mathbf{v%
}_{e1}\times \mathbf{B}_{0},
\end{equation}%
\begin{equation}
\frac{\partial n_{e1}}{\partial t}+v_{e1z}\frac{\partial n_{e0}}{\partial z}%
+n_{e0}\mathbf{\nabla }\cdot \mathbf{v}_{e1}=0,
\end{equation}%
\begin{equation}
\frac{\partial p_{e1}}{\partial t}+v_{e1z}\frac{\partial p_{e0}}{\partial z}%
+\gamma p_{e0}\mathbf{\nabla }\cdot \mathbf{v}_{e1}=-\left( \gamma -1\right)
\mathbf{\nabla \cdot q}_{e1},
\end{equation}%
\begin{equation}
\frac{\partial T_{e1}}{\partial t}+v_{e1z}\frac{\partial T_{e0}}{\partial z}%
+\left( \gamma -1\right) T_{e0}\mathbf{\nabla }\cdot \mathbf{v}_{e1}=-\left(
\gamma -1\right) \frac{1}{n_{e0}}\mathbf{\nabla \cdot q}_{e1},
\end{equation}%
\begin{equation}
\mathbf{q}_{e1}=-\mathbf{b}_{1}\chi _{e0}\frac{\partial T_{e0}}{\partial z}-%
\mathbf{b}_{0}\chi _{e0}\frac{\partial T_{e1}}{\partial z}-\mathbf{b}%
_{0}\chi _{e1}\frac{\partial T_{e0}}{\partial z},
\end{equation}%
\begin{equation}
\mathbf{F}_{e1}=q_{e}\mathbf{E}_{1}-m_{e}\nu _{ei}\left( \mathbf{v}_{e1}-%
\mathbf{v}_{i1}\right) .
\end{equation}
Here, $\chi _{e1}=5\chi _{e0}T_{e1}/2T_{e0}$ (and $\chi _{e}\sim T_{e}^{5/2}$%
, see Spitzer (1962)) is the perturbation of the thermal flux conductivity
coefficient. The perturbation of the unit magnetic vector $\mathbf{b}_{1}$
is equal to $b_{1x,y}=B_{1x,y}/B_{0}$ and $b_{1z}=0$. The thermal flux in
equilibrium is $\mathbf{q}_{e0}=-\mathbf{b}_{0}\chi _{e0}\frac{\partial
T_{e0}}{\partial z}$.

We have seen above at consideration of the ion perturbations that the terms $%
\sim 1/H^{2}$, where $H$ is the typical scale height, are needed to be kept
(see the last term in Equation (40)). Therefore, these terms are kept also
for the electrons.

\subsection{Equation for the electron temperature perturbation}

Let us find equation for the electron temperature perturbation. The
expression  $\mathbf{\nabla \cdot q}_{e1}$, where $\mathbf{q}_{e1}$ is
defined by (47), is given by%
\begin{equation}
\mathbf{\nabla \cdot q}_{e1}=\frac{\partial q_{e1y}}{\partial y}+\frac{%
\partial q_{e1z}}{\partial z}=-\chi _{e0}\frac{\partial T_{e0}}{\partial z}%
\frac{1}{B_{0}}\frac{\partial B_{1y}}{\partial y}-\chi _{e0}\frac{\partial
^{2}T_{e1}}{\partial z^{2}}-2\frac{\partial \chi _{e0}}{\partial z}\frac{%
\partial T_{e1}}{\partial z}-\frac{\partial ^{2}\chi _{e0}}{\partial z^{2}}%
T_{e1}.
\end{equation}
Substituting this expression into Equation (46), we obtain%
\begin{equation}
D_{1}T_{e1}=-v_{e1z}\frac{\partial T_{e0}}{\partial z}-\left( \gamma
-1\right) T_{e0}\mathbf{\nabla }\cdot \mathbf{v}_{e1}+\left( \gamma
-1\right) \frac{\chi _{e0}}{n_{e0}}\frac{\partial T_{e0}}{\partial z}\frac{%
\partial B_{1y}}{B_{0}\partial y},
\end{equation}
where the operator $D_{1}$ is defined as%
\begin{equation}
D_{1}=\left[ \frac{\partial }{\partial t}-\left( \gamma -1\right) \frac{1}{%
n_{e0}}\left( \chi _{e0}\frac{\partial ^{2}}{\partial z^{2}}+2\frac{\partial
\chi _{e0}}{\partial z}\frac{\partial }{\partial z}+\frac{\partial ^{2}\chi
_{e0}}{\partial z^{2}}\right) \right] .
\end{equation}

\subsection{Perturbed velocity and temperature of electrons}

We find now equations for components of the perturbed velocity of electrons.
The $x$-component of Equation (43) has a simple form, i.e.
\begin{equation}
v_{e1y}=-\frac{1}{m_{e}\omega _{ce}}F_{e1x},
\end{equation}
where $\omega _{ce}=q_{e}B_{0}/m_{e}c$. Applying the operator $\partial
/\partial t$ to the $y$-component of Equation (43) and using Equations (45)
and (49), we obtain%
\begin{eqnarray}
\frac{\partial }{\partial t}\left( v_{e1x}-\frac{1}{m_{e}\omega _{ce}}%
F_{e1y}\right) &\mathbf{=}&\mathbf{-}\frac{1}{\omega _{ci}}\frac{\partial
P_{e1}}{\partial y}-\left( \gamma -1\right) \frac{\chi _{e0}}{m_{e}\omega
_{ce}n_{e0}}\frac{\partial T_{e0}}{\partial z}\frac{\partial ^{2}B_{1y}}{%
B_{0}\partial y^{2}} \\
&&+\frac{1}{m_{e}\omega _{ce}}\left( D_{1}-\frac{\partial }{\partial t}%
\right) \frac{\partial T_{e1}}{\partial y},  \nonumber
\end{eqnarray}
where
\begin{equation}
P_{e1}=-g_{e}v_{e1z}+c_{se}^{2}\mathbf{\nabla }\cdot \mathbf{v}_{e1}
\end{equation}
and $c_{se}^{2}=\gamma p_{e0}/$ $m_{i}n_{e0}$. The variable $P_{e1}$ is
analogous to $P_{i1}$ (see Eq. [22]), which defines the ion pressure
perturbation. But for electrons, their pressure perturbation is also
affected by the thermal conductivity (see Eq. [45]).

Let us express $\mathbf{\nabla }\cdot \mathbf{v}_{e1}$ through $v_{e1z}$,
using Equation (52),%
\begin{equation}
\mathbf{\nabla }\cdot \mathbf{v}_{e1}=\frac{\partial v_{e1z}}{\partial z}-%
\frac{1}{m_{e}\omega _{ce}}\frac{\partial F_{e1x}}{\partial y}.
\end{equation}
The $z$-component of Equation (43) takes the form
\begin{equation}
0\mathbf{=-}\frac{1}{n_{e0}}\frac{\partial p_{e1}}{\partial z}+\frac{1}{%
n_{e0}}\frac{\partial p_{e0}}{\partial z}\frac{n_{e1}}{n_{e0}}+F_{e1z}.
\end{equation}

We consider further perturbations with the dynamic frequency $\partial
/\partial t$ satisfying the following conditions:%
\begin{equation}
\frac{\chi _{e0}}{n_{e0}}\frac{\partial ^{2}}{\partial z^{2}}\gg \frac{%
\partial }{\partial t}\gg \frac{1}{n_{e0}}\frac{\partial \chi _{e0}}{%
\partial z}\frac{\partial }{\partial z}.
\end{equation}

In this case, the terms proportional to $\partial \chi _{e0}/\partial z$ in
the temperature equation (50) (see [51]) are unimportant because the
necessary small corrections proportional to $\partial /\partial t$ in this
equation will be larger than that $\sim \partial \chi _{e0}/\partial z$.
Thus, an inhomogeneity of the thermal flux conductivity coefficient and its
perturbation can be neglected. We further apply the operator $\partial
/\partial t$ to Equation (56) and use Equations (44), (45), (49), and (55).
As a result, we obtain%
\begin{eqnarray}
\left( c_{se}^{2}\frac{\partial }{\partial z}-\gamma g_{e}\right) \frac{%
\partial v_{e1z}}{\partial z} &\mathbf{=}&\mathbf{-}\frac{\partial F_{e1z}}{%
m_{i}\partial t}+\left[ \left( 1-\gamma \right) g_{e}+c_{se}^{2}\frac{%
\partial }{\partial z}\right] \frac{1}{m_{e}\omega _{ce}}\frac{\partial
F_{e1x}}{\partial y} \\
&&+\left( \gamma -1\right) \frac{\chi _{e0}}{m_{i}n_{e0}}\left( \frac{%
\partial T_{e0}}{\partial z}\frac{1}{B_{0}}\frac{\partial ^{2}B_{1y}}{%
\partial y\partial z}+\frac{\partial ^{3}T_{e1}}{\partial z^{3}}\right) .
\nonumber
\end{eqnarray}

Equation for the temperature perturbation under conditions (57) has the form

\begin{eqnarray}
\left[ \left( \gamma -1\right) \frac{\chi _{e0}}{n_{e0}}\frac{\partial ^{2}}{%
\partial z^{2}}-\frac{\partial }{\partial t}\right] T_{e1} &=&v_{e1z}\frac{%
\partial T_{e0}}{\partial z}+\left( \gamma -1\right) T_{e0}\left( \frac{%
\partial v_{e1z}}{\partial z}-\frac{1}{m_{e}\omega _{ce}}\frac{\partial
F_{e1x}}{\partial y}\right) \\
&&-\left( \gamma -1\right) \frac{\chi _{e0}}{n_{e0}}\frac{\partial T_{e0}}{%
\partial z}\frac{\partial B_{1y}}{B_{0}\partial y},  \nonumber
\end{eqnarray}
where we have used Equation (55). Substituting $T_{e1}$ in Equation (58) and
carrying out some transformations, we find equation for the longitudinal
velocity $v_{e1z}$%
\begin{eqnarray}
\frac{\partial ^{3}v_{e1z}}{\partial z^{3}} &=&\mathbf{-}\frac{\partial
^{2}F_{e1z}}{T_{e0}\partial z\partial t}\mathbf{-}\frac{n_{e0}}{\chi _{e0}}%
\left( \frac{\partial }{\partial z}\right) ^{-1}\frac{\partial ^{2}F_{e1z}}{%
T_{e0}\partial t^{2}}+\frac{1}{m_{e}\omega _{ce}}\frac{\partial ^{3}F_{e1x}}{%
\partial y\partial z^{2}} \\
&&+\frac{1}{c_{se}^{2}}\left( \gamma g_{e}+\frac{\partial c_{se}^{2}}{%
\partial z}\right) \frac{1}{m_{e}\omega _{ce}}\frac{\partial ^{2}F_{e1x}}{%
\partial y\partial z}-\frac{\partial T_{e0}}{T_{e0}\partial z}\frac{1}{B_{0}}%
\frac{\partial ^{2}B_{1y}}{\partial y\partial t}.  \nonumber
\end{eqnarray}
The correction proportional to $\partial F_{e1x}/\partial t$ is absent. The
last term on the right hand-side of Equation (60) is connected with the
background electron thermal flux (Quataert 2008).

From Equations (59) and (60), we can find equation for the temperature
perturbation
\begin{eqnarray}
\left( \gamma -1\right) \frac{\chi _{e0}}{n_{e0}}\frac{\partial }{\partial z}%
\left( \frac{\partial ^{2}T_{e1}}{\partial z^{2}}+\frac{\partial T_{e0}}{%
\partial z}\frac{\partial B_{1y}}{B_{0}\partial y}\right) &=&\frac{\gamma
T_{e0}}{c_{se}^{2}}\left[ \left( \gamma -1\right) g_{e}+\frac{\partial
c_{se}^{2}}{\partial z}\right] \frac{1}{m_{e}\omega _{ce}}\frac{\partial
F_{e1x}}{\partial y} \\
&&-\left( \gamma -1\right) \frac{\partial F_{e1z}}{\partial t}-\gamma \frac{%
n_{e0}}{\chi _{e0}}\left( \frac{\partial }{\partial z}\right) ^{-2}\frac{%
\partial ^{2}F_{e1z}}{\partial t^{2}}  \nonumber \\
&&-\gamma \frac{\partial T_{e0}}{\partial z}\left( \frac{\partial }{\partial
z}\right) ^{-1}\frac{\partial ^{2}B_{1y}}{B_{0}\partial y\partial t}.
\nonumber
\end{eqnarray}
It is followed from results obtained below that all terms on the right-hand
side of Equation (61) (except the correction $\sim \partial
^{2}F_{e1z}/\partial t^{2}$) have the same order of magnitude (see Section
[4.3]). The left-hand side of this equation is larger (see conditions [57]).
Thus, the temperature perturbation in the zero order of magnitude can be
found by equaling the left part of Equation (61) to zero. However, the right
side is necessary for finding the transverse velocity perturbation $v_{e1x}$%
%. \

To find the velocity $v_{e1x}$, we need to calculate the value $P_{e1}$ (see
Eqs. [53] and [54]). Performing calculations in the same way as that for
ions (see Section 4.3), we obtain%
\begin{eqnarray}
c_{se}^{2}\frac{\partial ^{2}P_{e1}}{\partial z^{2}} &=&\left[ c_{se}^{2}%
\frac{\partial }{\partial z}+\left( \gamma -1\right) g_{e}+\frac{\partial
c_{se}^{2}}{\partial z}\right] \left( \mathbf{-}\frac{\partial F_{e1z}}{%
m_{i}\partial t}+\frac{\partial V_{e1}}{\partial z}\right) \\
&&+g_{e}\left[ \left( \gamma -1\right) g_{e}+\frac{\partial c_{se}^{2}}{%
\partial z}\right] \frac{1}{m_{e}\omega _{ce}}\frac{\partial F_{e1x}}{%
\partial y},  \nonumber
\end{eqnarray}
where we have introduced the notation connected with the thermal flux,%
\begin{equation}
V_{e1}=\left( \gamma -1\right) \frac{\chi _{e0}}{m_{i}n_{e0}}\left( \frac{%
\partial T_{e0}}{\partial z}\frac{1}{B_{0}}\frac{\partial B_{1y}}{\partial y}%
+\frac{\partial ^{2}T_{e1}}{\partial z^{2}}\right) .
\end{equation}

Equation (62) can be re-written in the form, which is convenient for finding
the velocity $v_{e1x}$. Using Equation (61), we obtain%
\begin{eqnarray}
\frac{\partial ^{2}}{\partial z^{2}}\left( P_{e1}-V_{e1}\right) &=&-\frac{%
\partial ^{2}F_{e1z}}{m_{i}\partial z\partial t}-\frac{\gamma }{c_{se}^{2}}%
\left[ \left( \gamma -1\right) g_{e}+\frac{\partial c_{se}^{2}}{\partial z}%
\right] \frac{\partial F_{e1z}}{m_{i}\partial t} \\
&&+\frac{1}{c_{se}^{2}}\left[ \left( \gamma -1\right) g_{e}+\frac{\partial
c_{se}^{2}}{\partial z}\right] \left( \gamma g_{e}+\frac{\partial c_{se}^{2}%
}{\partial z}\right) \frac{1}{m_{e}\omega _{ce}}\frac{\partial F_{e1x}}{%
\partial y}  \nonumber \\
&&-\left[ \left( \gamma -1\right) g_{e}+\frac{\partial c_{se}^{2}}{\partial z%
}\right] \frac{\partial T_{e0}}{T_{e0}\partial z}\left( \frac{\partial }{%
\partial z}\right) ^{-1}\frac{\partial ^{2}B_{1y}}{B_{0}\partial y\partial t}%
.  \nonumber
\end{eqnarray}

It is easy to see that Equation (53) has the form%
\begin{equation}
\frac{\partial }{\partial t}\left( v_{e1x}-\frac{1}{m_{e}\omega _{ce}}%
F_{e1y}\right) \mathbf{=-}\frac{1}{\omega _{ci}}\frac{\partial }{\partial y}%
\left( P_{e1}-V_{e1}\right) .
\end{equation}

Thus, the main contribution of the flux described by Equation (63) does not
influence on the electron dynamics. Applying to Equation (65) the operator $%
\partial ^{2}/\partial z^{2}$ and using Equation (64), we find an equation
for the velocity $v_{e1x}$.

\bigskip

\section{FOURIER CURRENT COMPONENTS}

\subsection{Fourier velocity components of ions and electrons}

Let us give velocities of ions and electrons in the Fourier-representation.
From Equations (20), (21), and (40), we have%
\begin{equation}
v_{i1xk}\mathbf{=}\frac{1}{\omega _{ci}^{2}}\left( 1+\frac{\omega ^{2}}{%
\omega _{ci}^{2}}\right) Q_{i1xk}+i\frac{k_{y}^{2}}{k_{z}^{2}}\frac{\left(
\omega ^{2}-g_{i}a_{i}\right) }{\omega \omega _{ci}^{3}}Q_{i1yk}\mathbf{-}%
\frac{1}{\omega _{ci}}\frac{k_{y}}{k_{z}}\left( 1-i\frac{a_{i}}{k_{z}}%
\right) F_{i1zk},
\end{equation}%
\begin{equation}
v_{i1yk}\mathbf{=}\frac{1}{\omega _{ci}^{2}}\left[ 1+\frac{\left(
k^{2}\omega ^{2}-k_{y}^{2}g_{i}a_{i}\right) }{k_{z}^{2}\omega _{ci}^{2}}%
\right] Q_{i1yk}+i\frac{\omega }{\omega _{ci}^{2}}\frac{k_{y}}{k_{z}}\left(
1-i\frac{a_{i}}{k_{z}}\right) F_{i1zk}.
\end{equation}
Here and below, we have introduced notations%
\begin{equation}
a_{i,e}=\frac{1}{c_{si,e}^{2}}\left[ \left( \gamma -1\right) g_{i,e}+\frac{%
\partial c_{si,e}^{2}}{\partial z}\right] .
\end{equation}
The velocity $v_{i1zk}$ is given by Equation (38).

From Equations (64) and (65), we find%
\begin{eqnarray}
v_{e1xk} &\mathbf{=}&\mathbf{-}i\frac{a_{e}c_{se}^{2}}{\omega \omega _{ci}}%
\frac{k_{y}^{2}}{k_{z}^{2}}\left( b_{e}\frac{1}{m_{e}\omega _{ce}}%
F_{e1xk}+\omega \frac{\partial T_{e0}}{k_{z}T_{e0}\partial z}\frac{B_{1yk}}{%
B_{0}}\right) \\
&&+\frac{1}{m_{e}\omega _{ce}}F_{e1yk}\mathbf{-}\frac{k_{y}}{k_{z}}\left(
1-i\gamma \frac{a_{e}}{k_{z}}\right) \frac{1}{m_{e}\omega _{ce}}F_{e1zk},
\nonumber
\end{eqnarray}
where the following notation is introduced:
\begin{equation}
b_{e}=\frac{1}{c_{se}^{2}}\left( \gamma g_{e}+\frac{\partial c_{se}^{2}}{%
\partial z}\right) .
\end{equation}
Equation (60) also gives us
\begin{eqnarray}
v_{e1zk} &=&\frac{k_{y}}{k_{z}}\frac{1}{m_{e}\omega _{ce}}F_{e1xk}-i\frac{%
k_{y}}{k_{z}^{2}}\left( b_{e}\frac{1}{m_{e}\omega _{ce}}F_{e1xk}+\omega
\frac{\partial T_{e0}}{k_{z}T_{e0}\partial z}\frac{B_{1yk}}{B_{0}}\right) \\
&&-i\frac{\omega }{k_{z}^{2}T_{e0}}\left( 1+i\omega \frac{n_{e0}}{\chi
_{e0}k_{z}^{2}}\right) F_{e1zk}.  \nonumber
\end{eqnarray}
The velocity $v_{e1y}$ is defined by Equation (52).

\subsection{Fourier electron velocity components at the absence of heat flux}

To elucidate the role of the electron thermal flux, we also consider the
dispersion relation when the flux is absent. Therefore, we give here the
corresponding electron velocity components:
\begin{equation}
v_{e1xk}=-i\frac{k_{y}^{2}g_{e}a_{e}}{k_{z}^{2}\omega \omega _{ci}}\frac{1}{%
m_{e}\omega _{ce}}F_{e1xk}+\frac{1}{m_{e}\omega _{ce}}F_{e1yk}-\frac{k_{y}}{%
k_{z}}\left( 1-i\frac{a_{e}}{k_{z}}\right) \frac{1}{m_{e}\omega _{ce}}%
F_{e1zk},
\end{equation}%
\begin{equation}
v_{e1zk}\mathbf{=}\frac{k_{y}}{k_{z}}\left( 1-i\frac{g_{e}}{k_{z}c_{se}^{2}}%
\right) \frac{1}{m_{e}\omega _{ce}}F_{e1xk}\mathbf{-}i\frac{\omega }{%
k_{z}^{2}c_{se}^{2}m_{i}}\left( 1-i\frac{\gamma g_{e}}{k_{z}c_{se}^{2}}%
\right) F_{e1zk}.
\end{equation}
Comparing expressions (69) and (71) with these equations, we see that the
thermal flux under conditions (57) essentially modifies the small terms in
the electron velocity.

\subsection{Fourier components of the current}

We find now the Fourier components of the linear current $\mathbf{j}%
_{1}=q_{i}n_{i0}\mathbf{v}_{i1}+q_{e}n_{e0}\mathbf{v}_{e1}$. It is
convenient to consider the value $4\pi i\mathbf{j}_{1}/\omega $. Using
expressions (38), (52), and (66)-(71), we obtain the following current
components:%
\begin{eqnarray}
\frac{4\pi i}{\omega }j_{1xk} &=&a_{xx}E_{1xk}+ia_{xy}E_{1yk}-a_{xz}E_{1zk}
\\
&&-b_{xx}\left( v_{i1xk}-v_{e1xk}\right) -ib_{xy}\left(
v_{i1yk}-v_{e1yk}\right) +b_{xz}\left( v_{i1zk}-v_{e1zk}\right) ,  \nonumber
\end{eqnarray}%
\ \
\begin{eqnarray}
\frac{4\pi i}{\omega }j_{1yk} &=&-ia_{yx}E_{1xk}+a_{yy}E_{1yk}-a_{yz}E_{1zk}
\\
&&+ib_{yx}\left( v_{i1xk}-v_{e1xk}\right) -b_{yy}\left(
v_{i1yk}-v_{e1yk}\right) +b_{yz}\left( v_{i1zk}-v_{e1zk}\right) ,  \nonumber
\end{eqnarray}%
\begin{eqnarray}
\frac{4\pi i}{\omega }j_{1zk} &=&-a_{zx}E_{1xk}-a_{zy}E_{1yk}+a_{zz}E_{1z} \\
&&+b_{zx}\left( v_{i1x}-v_{e1x}\right) +b_{zy}\left( v_{i1y}-v_{e1y}\right)
-b_{zz}\left( v_{i1z}-v_{e1z}\right) .  \nonumber
\end{eqnarray}

When obtaining expressions (74)-(76), we have used notations (16), (23),
(24), and (48) and equalities $q_{e}=-q_{i}$, $n_{e0}=n_{i0}$, $m_{e}\nu
_{ei}=m_{i}\nu _{ie}$. We have also substituted $B_{1yk}$ by $(k_{z}c/\omega
)E_{1xk}$ (see below). The following notations are introduced above:%
\begin{eqnarray}
a_{xx} &=&\frac{\omega _{pi}^{2}}{\omega _{ci}^{2}}\frac{k^{2}}{k_{z}^{2}}%
\left( 1-\frac{k_{y}^{2}}{k^{2}}\frac{g_{i}a_{i}+a_{e}b_{e}c_{se}^{2}}{%
\omega ^{2}}-\frac{k_{y}^{2}}{k^{2}}\frac{a_{e}c_{se}^{2}}{\omega ^{2}}\frac{%
\partial T_{e0}^{\ast }}{T_{e0}\partial z}\right) , \\
a_{xy} &=&a_{yx}=\frac{\omega _{pi}^{2}\omega }{\omega _{ci}^{3}}\frac{k^{2}%
}{k_{z}^{2}}\left( 1-\frac{k_{y}^{2}}{k^{2}}\frac{g_{i}a_{i}}{\omega ^{2}}%
\right) ,a_{xz}=\frac{\omega _{pi}^{2}}{\omega \omega _{ci}}\frac{k_{y}}{%
k_{z}^{2}}\left( a_{i}-\gamma a_{e}\right) ,  \nonumber \\
a_{yy} &=&\frac{\omega _{pi}^{2}}{\omega _{ci}^{2}},a_{yz}=a_{zy}=\frac{%
\omega _{pi}^{2}}{\omega _{ci}^{2}}\frac{k_{y}}{k_{z}},a_{zx}=\frac{\omega
_{pi}^{2}}{\omega \omega _{ci}}\frac{k_{y}}{k_{z}^{2}}\left( b_{e}-\frac{%
g_{i}}{c_{si}^{2}}+\frac{\partial T_{e0}^{\ast }}{T_{e0}\partial z}\right) ,
\nonumber \\
a_{zz} &=&\frac{\omega _{pi}^{2}}{k_{z}^{2}}\left( \frac{\gamma }{c_{se}^{2}}%
+\frac{1}{c_{si}^{2}}\right)  \nonumber
\end{eqnarray}
and%
\begin{eqnarray}
b_{xx} &=&\frac{\omega _{pi}^{2}\nu _{ie}}{\omega _{ci}^{2}}\frac{m_{i}}{%
q_{i}}\frac{k^{2}}{k_{z}^{2}}\left( 1-\frac{k_{y}^{2}}{k^{2}}\frac{%
g_{i}a_{i}+a_{e}c_{se}^{2}b_{e}}{\omega ^{2}}\right) , \\
b_{zx} &=&\frac{\omega _{pi}^{2}}{\omega \omega _{ci}}\frac{k_{y}}{k_{z}^{2}}%
\left( b_{e}-\frac{g_{i}}{c_{si}^{2}}\right) \frac{m_{i}}{q_{i}}\nu _{ie},
\nonumber \\
b_{ij} &=&a_{ij}\frac{m_{i}}{q_{i}}\nu _{ie}.  \nonumber
\end{eqnarray}
Here $\omega _{pi}=\left( 4\pi n_{i0}q_{i}^{2}/m_{i}\right) ^{1/2}$ is the
plasma frequency and $k^{2}=k_{y}^{2}+k_{z}^{2}$. The terms proportional to $%
T_{e0}^{\ast }$ are connected with the background electron thermal flux.

Calculations show that to obtain expressions for $a_{ij}$ without thermal
flux, using electron velocities (72) and (73), we must change $b_{e}$ by $%
g_{e}/c_{se}^{2}$, put $T_{e0}^{\ast }=0$, and take $\gamma =1$ in terms $%
a_{xz}$ and $a_{zz}$.

\subsection{Simplification of collision contribution}

From the formal point of view, an assumption that electrons are magnetized
has only been involved in neglecting the transverse electron thermal flux.
In other respects, a relationship between $\omega _{ce}$ and $\nu _{ei}$ or $%
\omega _{ci}$ and $\nu _{ie}$ (that is the same) can be arbitrary in
Equations (74)-(76). We further proceed by assuming that $\omega \ll \omega
_{ci}$. In this case, we can neglect the collisional terms proportional to $%
b_{xy}$ and $b_{yx}$. However, the system of Equations (74)-(76) stays
sufficiently complex to find $\mathbf{j}_{1}$ through $\mathbf{E}_{1}$.
Therefore, we further consider the specific case in which the frequency $%
\omega $ and wave numbers satisfy the following conditions:%
\begin{equation}
\frac{\omega _{ci}^{2}}{\nu _{ie}^{2}}\frac{k_{z}^{2}}{k^{2}}\gg \frac{%
\omega }{\nu _{ie}}\gg \frac{1}{k_{z}^{2}H^{2}}\frac{k_{y}^{2}c_{s}^{2}}{%
\omega _{ci}^{2}},
\end{equation}
where%
\begin{equation}
c_{s}^{2}=\frac{c_{si}^{2}c_{se}^{2}}{\gamma c_{si}^{2}+c_{se}^{2}}.
\end{equation}

It is clear that conditions (79) can easily be realized. In this case, the
current components are equal to%
\begin{eqnarray}
\frac{4\pi i}{\omega }j_{1xk} &=&\varepsilon _{xx}E_{1xk}+i\varepsilon
_{xy}E_{1yk}-\varepsilon _{xz}E_{1zk}, \\
\frac{4\pi i}{\omega }j_{1yk} &=&-i\varepsilon _{yx}E_{1xk}+\varepsilon
_{yy}E_{1yk}-\varepsilon _{yz}E_{1zk},  \nonumber \\
\frac{4\pi i}{\omega }j_{1zk} &=&-\varepsilon _{zx}E_{1xk}-\varepsilon
_{zy}E_{1yk}+\varepsilon _{zz}E_{1z}.  \nonumber
\end{eqnarray}

Components of the dielectric permeability tensor $\varepsilon _{ij}$ are the
following:
\begin{eqnarray}
\varepsilon _{xx} &=&a_{xx}+i\frac{\nu _{ie}}{\omega _{ci}}\frac{k_{y}}{%
k_{z}^{2}}\frac{\left( a_{i}-\gamma a_{e}\right) }{\left( 1-id_{z}\right) }%
a_{zx},\varepsilon _{xy}=a_{xy}+\frac{\nu _{ie}}{\omega _{ci}}\frac{k_{y}}{%
k_{z}^{2}}\frac{\left( a_{i}-\gamma a_{e}\right) }{\left( 1-id_{z}\right) }%
a_{zy}, \\
\varepsilon _{xz} &=&\frac{a_{xz}}{\left( 1-id_{z}\right) },\varepsilon
_{yx}=a_{yx}-\frac{\omega \nu _{ie}}{\omega _{ci}^{2}}\frac{k_{y}}{k_{z}}%
\frac{a_{zx}}{\left( 1-id_{z}\right) },\varepsilon _{yy}=a_{yy},  \nonumber
\\
\varepsilon _{yz} &=&\frac{a_{yz}}{\left( 1-id_{z}\right) },\varepsilon
_{zx}=\frac{a_{zx}}{\left( 1-id_{z}\right) },\varepsilon _{zy}=\frac{a_{zy}}{%
\left( 1-id_{z}\right) },\varepsilon _{zz}=\frac{a_{zz}}{\left(
1-id_{z}\right) },  \nonumber
\end{eqnarray}
where we have used notations (78)
\begin{equation}
d_{z}=\frac{\omega \nu _{ie}}{k_{z}^{2}c_{s}^{2}}.
\end{equation}
Parameter $d_{z}$ defines the collisionless, $d_{z}\ll 1$, and collisional, $%
d_{z}\gg 1$ regimes. Below, we derive the dispersion relation.

\bigskip

\section{DISPERSION\ RELATION}

From Equations (9) and (10) in the Fourier-representation and using system
of equations (81), we obtain the following equations for the electric field
components:
\begin{eqnarray}
\left( n^{2}-\varepsilon _{xx}\right) E_{1xk}-i\varepsilon
_{xy}E_{1yk}+\varepsilon _{xz}E_{1zk} &=&0, \\
i\varepsilon _{yx}E_{1xk}+\left( n_{z}^{2}-\varepsilon _{yy}\right)
E_{1yk}+\left( -n_{y}n_{z}+\varepsilon _{yz}\right) E_{1zk} &=&0,  \nonumber
\\
\varepsilon _{zx}E_{1xk}+\left( -n_{y}n_{z}+\varepsilon _{zy}\right)
E_{1yk}+\left( n_{y}^{2}-\varepsilon _{zz}\right) E_{1zk} &=&0,  \nonumber
\end{eqnarray}
where $\mathbf{n=k}c/\omega $. The dispersion relation can be found by
setting the determinant of the system (84) equal to zero. In our case, the
terms proportional to $\varepsilon _{xy}$ and $\varepsilon _{yx}$ can be
neglected. As a result, we have%
\[
\left( n^{2}-\varepsilon _{xx}\right) \left[ n_{y}^{2}\varepsilon
_{yy}+\left( n_{z}^{2}-\varepsilon _{yy}\right) \varepsilon
_{zz}-n_{y}n_{z}\left( \varepsilon _{yz}+\varepsilon _{zy}\right)
+\varepsilon _{yz}\varepsilon _{zy}\right]
\]
\begin{equation}
+\left( n_{z}^{2}-\varepsilon _{yy}\right) \varepsilon _{xz}\varepsilon
_{zx}=0.
\end{equation}

The above dispersion relation can be studied for different cases. In
subsequent sections, we consider both the collisionless and collisional
cases.

\subsection{ Collisionless case}

We assume now that the condition%
\begin{equation}
\frac{\omega \nu _{ie}}{k_{z}^{2}c_{s}^{2}} \ll 1,
\end{equation}
is satisfied. Then, using notations (77) and (82), the dispersion relation
(85) becomes%
\begin{equation}
\left( \omega ^{2}-k_{z}^{2}c_{A}^{2}\right) \left( \omega
^{2}-k_{z}^{2}c_{A}^{2}-\Omega ^{2}\frac{k_{y}^{2}}{k^{2}}\right) =0,
\end{equation}
where $c_{A}=B_{0}/(4\pi m_{i}n_{i0})^{1/2}$ is the Alfv\'{e}n velocity and
\begin{equation}
\Omega ^{2}=g_{i}a_{i}+c_{se}^{2}a_{e}b_{e}+c_{se}^{2}a_{e}\frac{\partial
T_{e0}^{\ast }}{T_{e0}\partial z}+c_{s}^{2}\left( a_{i}-\gamma a_{e}\right)
\left( b_{e}-\frac{g_{i}}{c_{si}^{2}}+\frac{\partial T_{e0}^{\ast }}{%
T_{e0}\partial z}\right) .
\end{equation}

For obtaining Equation (87), we have used the condition $k_{y}^{2}c_{s}^{2}/%
\omega _{ci}^{2} \ll 1$. We see that there are two wave modes. The first
wave mode, $\omega ^{2}=k_{z}^{2}c_{A}^{2}$, is the Alfv\'{e}n wave with a
polarization of the electric field mainly along the $y$-axis (the wave
vector $\mathbf{k}$ is situated in the $y-z$ plane). This wave does not feel
the inhomogeneity of the medium. The second wave has a polarization of the
magnetosonic wave, i.e. its electric field is directed mainly along the $x$%
-axis (see below). This wave is undergone by the action of the medium
inhomogeneity effect. The corresponding dispersion relation is
\begin{equation}
\omega ^{2}=k_{z}^{2}c_{A}^{2}+\Omega ^{2}\frac{k_{y}^{2}}{k^{2}}.
\end{equation}

The expression (88) can be further simplified using equations (11), (12),
(68), (70), and (80). As a result, we obtain%
\begin{equation}
\Omega ^{2}=\frac{\gamma }{\gamma c_{si}^{2}+c_{se}^{2}}\frac{1}{m_{i}^{2}}%
\left[ \left( \gamma -1\right) m_{i}g+\gamma \frac{\partial \left(
T_{i0}+T_{e0}\right) }{\partial z}\right] \left[ m_{i}g+\frac{\partial
\left( T_{e0}+T_{e0}^{\ast }\right) }{\partial z}\right] .
\end{equation}%
We have pointed out at the end of Section (6.3) what changes must be
done in expressions (77) and (78) to consider the case without heat flux.
This case follows from Equation (90), if we omit the term $\partial \left(
T_{e0}+T_{e0}^{\ast }\right) /\partial z$ and put $\gamma =1$
in the first multiplier. Then $\Omega ^{2}$ becomes 

\begin{equation}
\Omega ^{2}=\frac{g}{c_{si}^{2}+c_{se}^{2}}\left[ \left( \gamma -1\right) g+%
\frac{\partial \left( c_{si}^{2}+c_{se}^{2}\right) }{\partial z}\right] .
\end{equation}

This is the Brunt-V\"{a}is\"{a}l\"{a} frequency. Comparing (90) and (91), we
see that the heat flux stabilizes the unstable stratification. The presence
of the background heat flux does not play of principle role. If the
temperature decreases in the direction of gravity ($\partial
T_{i,e0}/\partial z>0$), a medium is stable. Solution (90) describes an
instability regime only when
\[
\frac{\gamma -1}{2\gamma }m_{i}g<-\frac{\partial T_{0}}{\partial z}<\frac{1}{%
2}m_{i}g.
\]%
where ($T_{i0}\sim T_{e0}=T_{0}$). We also note that $\Omega ^{2}$ can be
negative if gradients of $T_{i0}$ and $T_{e0}$ have different signs.

For a comparison, we give here the corresponding dispersion relation
by Quataert (2008)%
\[
\omega ^{2}\simeq -g\left( \frac{d\ln T_{0}}{dz}\right) \frac{k_{\perp }^{2}%
}{k^{2}},
\]
which is discussed in Section 8.

\subsection{Collisional case}

We proceed with the collisional case when
\begin{equation}
\frac{\omega \nu _{ie}}{k_{z}^{2}c_{s}^{2}}\gg 1.
\end{equation}
In this limiting case, we obtain again Equation (89).

\subsection{Polarization of perturbations}

Let us neglect in the system of equations (84) the small contributions given
by $\varepsilon _{xy}$ and $\varepsilon _{yx}$. Then, for example, in the
collisionless case, we obtain for the second wave $\omega ^{2}\neq
k_{z}^{2}c_{A}^{2}$,%
\begin{eqnarray}
E_{1yk} &=&\frac{k_{y}}{k_{z}}E_{1zk}, \\
E_{1zk} &=&\frac{\varepsilon _{zx}}{\varepsilon _{zz}}E_{1xk}\ll E_{1xk}.
\nonumber
\end{eqnarray}

Thus, the second wave has a polarization of the electric field mainly along
the $x$-axis. In spite of that the component $E_{1zk}\ll E_{1xk}$, it is
multiplied by a large coefficient in the first equation of the system (84).
As a result, the contribution of this term is the same on the order of
magnitude as that of the first term.

In the collisional case, the component $E_{1zk}$ is also defined by Equation
(93). However, its contribution to the first equation of the system (84) can
be neglected.

\bigskip

\section{DISCUSSION}

Dispersion relation (87) with $\Omega ^{2}$ defined by Equations (88) or
(90) considerably differs from that given in (Quataert 2008) for the case of
our geometry. The reason goes back to the assumptions made in the MHD
analysis of buoyancy instabilities $p_{1}/p_{0}\ll \rho _{1}/\rho _{0}$,
where $p$ and $\rho $ denote the pressure and mass density of fluid, and the
condition of incompressibility $\mathbf{\nabla \cdot v}_{1}=0$, where $%
\mathbf{v}_{1}=\mathbf{v}_{i1}$ is the perturbed fluid velocity. We now
shortly show how one can obtain the result of Quataert (2008) in our
geometry, using these assumptions. We sum Equations (13) and (43) and use
the Ampere's law (10). The components of the equations become,
\begin{eqnarray}
\frac{\partial v_{i1x}}{\partial t} &\mathbf{=}&\frac{B_{0}}{4\pi \rho _{0}}%
\frac{\partial B_{1x}}{\partial z}, \\
\frac{\partial v_{i1y}}{\partial t} &\mathbf{=}&\mathbf{-}\frac{\partial
p_{1}}{\rho _{0}\partial y}-\frac{B_{0}}{4\pi \rho _{0}}\left( \frac{%
\partial B_{1z}}{\partial y}-\frac{\partial B_{1y}}{\partial z}\right) ,
\nonumber \\
\frac{\partial v_{i1z}}{\partial t} &\mathbf{=}&\mathbf{-}\frac{\partial
p_{1}}{\rho _{0}\partial y}-g\frac{n_{e1}}{n_{0}},  \nonumber
\end{eqnarray}
where $n_{i0}=n_{e0}=n_{0}$, $\rho _{0}=m_{i}n_{0}$, and we have used $%
n_{i1}=n_{e1}$. The components of the ideal magnetic induction equation are
the following:%
\begin{eqnarray}
\frac{\partial B_{1x}}{\partial t} &\mathbf{=}&B_{0}\frac{\partial v_{i1x}}{%
\partial z}\mathbf{,} \\
\frac{\partial B_{1y}}{\partial t} &\mathbf{=}&B_{0}\frac{\partial v_{i1y}}{%
\partial z},  \nonumber \\
\frac{\partial B_{1z}}{\partial t} &\mathbf{=}&-B_{0}\frac{\partial v_{i1y}}{%
\partial y}.  \nonumber
\end{eqnarray}

We note that the last equation (95) is a consequence of the second equation
(95) and $\mathbf{\nabla \cdot B}_{1}=0$.

The first equations of the systems of equations (94) and (95) describe the
Alfv\'{e}n waves, which are split from the other perturbations (see Eq.
[87]). Applying operators $\partial ^{3}/\partial z^{2}\partial t$ and $%
\partial ^{3}/\partial y\partial z\partial t$ to the second and third
equations of the systems of equations (94), correspondingly, using equation $%
\mathbf{\nabla \cdot v}_{i1}=0$ and the second equation (95), and
subtracting one equation from another, we obtain%
\begin{equation}
\left( \frac{\partial ^{2}}{\partial y^{2}}+\frac{\partial ^{2}}{\partial
z^{2}}\right) \frac{\partial ^{2}v_{i1y}}{\partial t^{2}}\mathbf{=}%
c_{A}^{2}\left( \frac{\partial ^{2}}{\partial y^{2}}+\frac{\partial ^{2}}{%
\partial z^{2}}\right) \frac{\partial ^{2}v_{i1y}}{\partial z^{2}}+\frac{g}{%
n_{0}}\frac{\partial ^{3}n_{e1}}{\partial y\partial z\partial t}.
\end{equation}
Also, from Equation (61), we have (see conditions [57]),
\begin{equation}
\frac{\partial ^{2}T_{e1}}{\partial z^{2}}=-\frac{\partial T_{e0}}{\partial z%
}\frac{\partial B_{1y}}{B_{0}\partial y}.
\end{equation}

Taking into account that $T_{e1}/T_{e0}=-n_{e1}/n_{e0}$, differentiating
Equation (97) over $t$, using the second equation (95), and substituting the
equation obtained in Equation (96), we find
\begin{equation}
\left( \frac{\partial ^{2}}{\partial y^{2}}+\frac{\partial ^{2}}{\partial
z^{2}}\right) \frac{\partial ^{2}v_{i1y}}{\partial t^{2}}\mathbf{=}%
c_{A}^{2}\left( \frac{\partial ^{2}}{\partial y^{2}}+\frac{\partial ^{2}}{%
\partial z^{2}}\right) \frac{\partial ^{2}v_{i1y}}{\partial z^{2}}+g\frac{%
\partial T_{e0}}{T_{e0}\partial z}\frac{\partial ^{2}v_{i1y}}{\partial y^{2}}%
.
\end{equation}
By neglecting the contribution of the magnetic field, this equation
coincides in the Fourier-representation with Equation (22) in Quataert
(2008).

However, the presence of the longitudinal electric field perturbations $%
E_{1z}$ results in that $p_{i,e1k}/p_{i,e0}\sim n_{i,e1k}/n_{i,e0}$ (for
ions, see Section [4.4]). Besides, Equation (97) together with equation $%
\mathbf{\nabla \cdot v}_{i1}=0$ lead to the nonphysical equation
\[
\frac{\partial T_{e1}}{\partial t}=v_{e1z}\frac{\partial T_{e0}}{\partial z}%
.
\]
Thus, Equation (98) is incorrect.

From the dispersion relation (85), we see the necessity of involving the
contribution of values $\varepsilon _{xz}$, $\varepsilon _{zx}$, and $%
\varepsilon _{zz}$ in the collisionless case (86) (values $%
\varepsilon _{xz}$ and $\varepsilon _{zx}$ give the last term on the right
hand-side of Eq. [88]). This means that contribution of currents $j_{1x}\sim
E_{1z}$ and $j_{1z}\sim E_{1x},E_{1z}$ must be taken into account. However,
the role of the longitudinal electric field $E_{1z}$ in the MHD equations is
not clear. The same also relates to the collisional case (92). In
the current $j_{1xk}$, we must take into consideration the
contribution of the current $j_{1zk}$ as a result of collisions,
which is proportional to $E_{1xk}$ (see Eqs. [74] and [76]).

Thus, the standard MHD equations with simplified assumptions are not
applicable for the correct theory of buoyancy instabilities. Such a theory
can only be given by the multicomponent approach used in this paper.

The results following from Equation (90) show that the thermal flux
stabilizes the buoyancy instability. The instability is only possible in the
narrow region of the temperature gradient (see Section [7.1]). The presence
of the background electron thermal heat (the term $\sim T_{e0}^{\ast }$)
does not play an essential role. An instability is also possibly, if the
temperature gradients of ions and electrons have the opposite signs.

The contribution of collisions between electrons and ions depends on the
parameter $d_{z}$ defined by Equation (83). In the both limits (86) ($%
d_{z}\ll 1$) and (92) ($d_{z}\gg 1$), the dispersion relation has the same
form.

We would like to say a few words about the Schwarzschild criterion of the
buoyancy instability. It is generally accepted that this instability is
possible, if the entropy increases in the direction of gravity. From a
formal point of view, it is correct, if we take the Brunt-V\"{a}is\"{a}l\"{a}
frequency $N$ in the form (e.g. Balbus 2000),%
\[
N^{2}=-\frac{1}{\gamma \rho }\frac{\partial p}{\partial z}\frac{\partial \ln
p\rho ^{-\gamma }}{\partial z}.
\]%
However, this expression can easily be transformed to expression (32). Thus,
we see that for instability to exist, the temperature must increase along
gravity and exceed the threshold.

\bigskip

\section{CONCLUSION}

In this paper, we have investigated buoyancy instabilities in magnetized
electron-ion astrophysical plasmas with the background electron thermal
flux, using the $\mathbf{E}$-approach when dynamical equations for the ions
and the electrons are solved separately via electric field perturbations. We
have included the background electron heat flux and collisions between
electrons and ions. The important role of the longitudinal electric field
perturbations, which are not captured by the MHD equations, has been shown.
We showed that the previous MHD\ result for the growth rate in the geometry
considered in this paper when all background quantities are directed along
the one axis is incorrect. The reason of this has been shown to be in
simplified assumptions made in the MHD analysis of the buoyancy
instabilities.

We have adopted that cyclotron frequencies of species are much larger than
their collision frequencies that is typical for ICM and galaxy clusters. The
dispersion relation obtained shows that the anisotropic electron heat flux,
including the background one, stabilizes the unstable stratification except
the narrow region of the temperature gradient. However, when gradients of
the ion and electron temperatures have opposite signs, the medium becomes
unstable.

Results obtained in this paper are applicable to the magnetized weakly
collisional stratified objects and can be useful for searching sources of
turbulent transport of energy and matter. It has been suggested that
buoyancy instability can act as a driving mechanism to generate turbulence
in ICM and this extra source of the heating may help to resolve cooling flow
problem. However, all previous analytical or numerical studies are
restricted to the MHD approach. Our study shows that when the true
multifluid nature of the system with the electron heat flux is considered,
one can not expect buoyancy instability unless for a very limited range of
the gradient of the temperature or when the gradients of the temperature of
the electrons and ions have opposite signs which both cases are very
unlikely. However, in the case when the heat flux does not play the role,
the system can be unstable according to the Schwarzschild criterion. 

The current linear analysis is for simplified initial conditions, in
which the background magnetic field, temperature gradient, and gravity are
along the same direction. However, another configuration should also be
examined using the $E$-approach, in which the initial magnetic field is
perpendicular to the direction of gravity. This will be done in the
forthcoming paper.

\bigskip

\subsection{References}

Balbus, S. A. 2000, ApJ, 534, 420

Balbus, S. A. 2001, ApJ, 562, 909

Balbus, S. A., \& Hawley, J. F. 1991, ApJ, 376, 214

Braginskii, S. I. 1965, Rev. Plasma Phys., 1, 205

Carilli, C. L., \& Taylor, G. B. 2002, ARA\&A, 40, 319

Chandran, B. D., \& Dennis, T. J. 2006, ApJ, 642, 140

Chang, P., \& Quataert, E. 2009, 0909.3041. (submitted to MNRAS)

Fabian, A. C., Sanders, J. S., Taylor, G. B., Allen, S. W., Crawford, C. S.,
Johnstone, R. M.,

\&\ Iwasawa, K. 2006, MNRAS, 366, 417

Gossard, E.E., \& Hooke, W.H. 1975, Waves in the Atmosphere (Amsterdam:
Elsevier Scientific Publishing Company)

Lyutikov, M. 2007, ApJL, 668, L1

Lyutikov, M. 2008, ApJL, 673, L115

Narayan, R., Igumenshchev, I. V., \& Abramowicz, M. A. 2000, ApJ, 539, 798

Narayan, R., Quataert, E., Igumenshchev, I. V., \& Abramowicz, M. A. 2002,
ApJ, 577, 295

Nekrasov, A. K. 1994 J. Atmos.Terr. Phys., 56, 931

Nekrasov, A. K. 2008, Phys. Plasmas, 15\textbf{,} 032907

Nekrasov, A. K. 2009 a, Phys.Plasmas, 16, 032902

Nekrasov, A. K. 2009 b, ApJ, 695, 46

Nekrasov, A. K. 2009 c, ApJ, 704, 80

Parrish, I. J., \& Quataert, E. 2008, ApJL, 677, L9

Parrish, I. J., Stone, J. M., \& Lemaster, N. 2008, ApJ, 688, 905

Parrish, I. J., Quataert, E., \& Sharma, P. 2009, ApJ, 703, 96

Pedlosky, J. 1982, Geophysical Fluid Dynamics, (New York: Springer-Verlag)

Quataert, E. 2008, ApJ, 673, 758

Rasera, Y., \& Chandran, B. 2008, ApJ, 685, 105

Ren, H., Wu, Z., Cao, J., \& Chu, P. K. 2009, Phys. Plasmas, 16, 102109

Sanders, J. S.,\ Fabian, A. C., Frank, K. A., Peterson, J. R., \& Russell,
H. R. 2010, MNRAS, 402, 127

Sarazin, C. L. 1988, X-Ray Emission from Clusters of Galaxies (Cambridge:

Cambridge Univ. Press)

Schwarzschild, M. 1958, Structure and Evolution of the Stars (New York:
Dover)

Sharma, P., Chandran, B. D. G., Quataert, E., \& Parrish, I. J. 2009, ApJ,
699, 348

Spitzer, L., Jr. 1962, Physics of Fully Ionized Gases (2d ed.; New York:
Interscience)

\end{document}